%Paper: hep-th/9508038
%From: Christopher Ford <cford@stp.dias.ie>
%Date: Wed, 9 Aug 1995 12:11:00 +0100 (BST)

\font\magnifiedtwelverm=cmr12 scaled\magstep3
\font\magnifiedtwelvebf=cmbx12 scaled\magstep3
\font\magnifiedtwelveit=cmti12 scaled\magstep3
\font\magnifiedtwelvesl=cmsl12 scaled\magstep3

\def\Bigtype{\let\rm=\magnifiedtwelverm \let\bf=\magnifiedtwelvebf
\let\it=\magnifiedtwelveit \let\sl=\magnifiedtwelvesl
\rm}

\font\twelverm=cmr12
\font\twelvebf=cmbx12
\font\twelveit=cmti12
\font\twelvesl=cmsl12

\def\bigtype{\let\rm=\twelverm \let\bf=\twelvebf
\let\it=\twelveit \let\sl=\twelvesl
\rm}

\def\ni{\noindent}
\def\qa{\quad}

\def\d{\partial}
\def\dq{\partial\cdot\partial}

\def\dm{\partial^\mu}
\def\dn{\partial^\nu}
\def\do{\partial^0}
\def\dM{\partial_\mu}
\def\dN{\partial_\nu}
\def\dO{\partial_0}
\def\dsm{\mathop{\dm}^{\leftrightarrow}}

\def\dso{\mathop{\do}^{\leftrightarrow}}
\def\dsO{\mathop{\dO}^{\leftrightarrow}}

\def\dsOx{\mathop{\d_{x^0}}^{\leftrightarrow}}
\def\dsOy{\mathop{\d_{y^0}}^{\leftrightarrow}}

\def\={\mathop{=}^{\rm def}}
\def\pM{{\scriptstyle{}^{\hphantom{(}+\hphantom{)}}_{(-)}}}

\def\intx{\int\limits_{x_0={\rm const.}} d^3\vec x}
\def\inty{\int\limits_{y_0={\rm const.}} d^3\vec y}
\def\intk{\int dk\,}

\def\Ama{A^\mu_a}   \def\AMa{A_{\mu a}}
\def\Ana{A^\nu_a}   \def\ANa{A_{\nu a}}
\def\Amb{A^\mu_b}   
\def\Anb{A^\nu_b}   \def\ANb{A_{\nu b}}
\def\Amc{A^\mu_c}   
\def\Anc{A^\nu_c}   

\def\Amap{(A_{\rm phys})^\mu_a}
\def\Anap{(A_{\rm phys})^\nu_a}

\def\Anbp{(A_{\rm phys})^\nu_b}
\def\Amau{(A_{\rm unphys})^\mu_a}

\def\jma{J^\mu_a}

\def\sma{S^\mu_a}

\def\vma{V^\mu_a}

\def\xma{X^\mu_a}

\def\Fmna{F^{\mu\nu}_a}    
    
\def\Fmnc{F^{\mu\nu}_c}    
\def\Gmna{G^{\mu\nu}_a}    
    
\def\Gmnc{G^{\mu\nu}_c}    

\def\tu{\tilde{u}}
\def\ua{u_a}  \def\ub{u_b}  \def\uc{u_c}
\def\tua{\tu_a}  \def\tub{\tu_b}  \def\tuc{\tu_c}

\def\ha{h_a} \def\hb{h_b} \def\hc{h_c}

\def\ka{k_a}  

\def\khi{\phi^+}
\def\Fi{\phi_i}    \def\Fj{\phi_j}    
\def\Ki{\khi_i}        
\def\osi{\overline{\psi}}
\def\Pi{\psi_i}    \def\Pj{\psi_j}    
\def\Qi{\osi_i}        

\def\f{f_{abc}}

\def\al{\alpha}
\def\be{\beta}
\def\ga{\gamma}
\def\de{\delta} \def\dab{\de_{ab}}
\def\ep{\epsilon}
\def\ka{\kappa}
\def\la{\lambda}

\def\zw{{1\over 2}}
\def\Tm{T^\mu}
\def\TM{T_{\rm matter}}
\def\TMm{T_{\rm matter}^\mu}
\def\TG{T_{\rm gauge}}
\def\TGm{T_{\rm gauge}^\mu}
\def\mas{(m^2_S)}
\def\mad{(m_D)}

\def\ei{e^{ikx}}    \def\emi{e^{-ikx}}

\def\Ua{\underline a}
\def\Ub{\underline b}
\def\Uc{\underline c}

\def\hp{H_{\rm phys}}
\def\hk{H_K}
\def\hr{H_R}
\def\hkk{H_{K^+}}
\def\hrk{H_{R^+}}

\def\pp{P_{\rm phys}}
\def\po{P_0}
\def\pk{P_K}
\def\pr{P_R}
\def\pkk{P_{K^+}}
\def\prk{P_{R^+}}

\def\Uao{\Ua_0}
\def\Ubo{\Ub_0}

\def\Uak{\Ua_K}
\def\Ubk{\Ub_K}
\def\Uar{\Ua_R}
\def\Ubr{\Ub_R}
\def\Uakk{\Ua_{K^+}}

\def\Uark{\Ua_{R^+}}
\def\Ubrk{\Ub_{R^+}}

\def\sp{S_{\rm phys}}
\def\ap{A_{\rm phys}}
\def\bp{B_{\rm phys}}

\def\p{{\rm phys}}

\def\ns{N^{\sim 1}}

\def\os{\,\oplus_\bot}
\def\om{\,\ominus_\bot}

\def\T{{\cal T}}

{\Bigtype {\bf A Causal Approach to Massive
\hfill{\bigtype DIAS-STP-95-01}
\vskip 0.3cm
Yang-Mills Theories }}
\vskip 1.5cm
{\bigtype F.Krahe}
\vskip 0.2cm
{\bigtype {\it Institute for Advanced Studies,
10 Burlington Road, Dublin 4, Ireland }}
\vskip 0.5cm
Work supported by Schweizerischer Nationalfonds
\vskip 1cm
{\bigtype {\bf Summary - }}We study quantized Yang-Mills theory with
massive vector fields in the framework of causal perturbation
theory. The most general form of the interaction which is invariant
under operator gauge transformations is pointed out. The generator of
these transformations generally fails to be nilpotent. This defect,
however, is easily cured by including scalar fields in the gauge
transformations. Due to gauge invariance these scalar gauge fields couple
to the Yang-Mills fields with predicted strength. We also show that invariance
under ghost charge conjugation fixes the form of the interaction completely.
The coupling of the Yang-Mills fields and the scalar gauge fields to matter is
investigated. It is proven that gauge invariance implies unitarity of the
physical $S$-matrix. We always work in the Fock space of free quantum fields
in which all expressions are mathematically well defined.

\vskip 1cm
{\bigtype {\bf 1. Introduction }}
\vskip 0.2cm

Recently, massless Yang-Mills theory has been succesfully studied in the
framework of causal perturbation theory [1-4]. The central object in
this approach is the causal $S$-matrix
$$S[g]=1+\sum_{n=1}^\infty\int d^4x_1\cdots d^4x_n
T^{(n)}(x_1,\cdots,x_n) \eqno(1.1)$$
$T^{(1)}$ specifies the theory.
For massless Yang Mills theories it is given by
$$T^{(1)}(x)\=-ief_{abc}\{\zw:\AMa\ANb\Fmnc:-:\AMa\ub\dm\tuc:\}(x)
\eqno(1.2)$$
$e$ is the coupling constant and $f_{abc}$ are the structure constants of
a non-abelian semi-simple compact gauge group $G$. $\Ama$ are the {\sl
free} gauge fields, defined by
$$\dq\Ama(x)=0, \quad [\Ama(x),\Anb(y)]_-=
i\delta_{ab}g^{\mu\nu}D_0(x-y) \eqno(1.3)$$
where $D_0$ is the Pauli - Jordan commutation function for $m=0$.
$\Fmna$ are the {\sl free} field strengths:
$$\Fmna\=\dm\Ana-\dn\Ama \eqno(1.4)$$
and $\ua$ and $\tua$ are the {\sl free} ghost fields:
$$\dq\ua(x)=\dq\tua(x)=0,\quad
\{\ua(x),\tub(y)\}_+=-i\delta_{ab}D_0(x-y) \eqno (1.5)$$
$$\{\ua(x),\ub(y)\}_+=\{\tua(x),\tub(y)\}_+=0 \eqno(1.6)$$
A detailed discussion of the algebraic properties of these ghost fields
can be found in [5].

Differentiating (1.3) we get
$$[\dM\Ama(x),\dN\Anb(y)]_-=0 \eqno(1.7)$$
$$[\dM\Ama(x),F^{\kappa\lambda}_b(y)]_-=0 \eqno(1.8)$$
Despite their simplicity, these equations have important consequences.
For, let us consider the operator
$$Q\=\intx\left(\dM\Ama(x)\right)\dsO \ua(x) \eqno (1.9)$$
Using the Leibnitz rule for graded algebras gives
$$\eqalignno{&Q^2=\zw\{Q,Q\}_+={1\over 2}\intx\inty
\Bigl\{[\dM\Ama(x),\dN\Anb(y)]_-\dsOx\dsOy\left(\ua(x)\ub(y)\right)+ \cr
&+\left(\dN\Anb(y)\dM\Ama(x)\right)\dsOx\dsOy\{\ua(x),\ub(y)\}_+\Bigr\}
=0&(1.10)\cr}$$

Thus eqs. (1.6, 1.7) make $Q$ a differential operator in the sense of
homological algebra. This allows for standard homological notions [6,7]:
Let ${\cal F}=\{F\}$ be the field algebra consisting of the polynomials in
the (smeared) gauge and ghost fields and their Wick powers. Consider
the ghost charge operator [5]
$$Q_g\=\intx\tua(x)\dsO\ua(x) \eqno (1.11)$$
and the corresponding derivation $\delta_g$ in ${\cal F}$
$$\delta_gF\=[Q_g,F]_- \eqno (1.12)$$
We say an operator F has ghost charge $z$ if
$$\delta_gF=zF \eqno (1.13)$$
Since $Q_g$ has integer spectrum [5] we have $z\in{\cal Z}$. The operators
$F_z$ with ghost charge $z$ form the subspace ${\cal F}_z$, and we
obviously have
$${\cal F}=\bigoplus_{z\in{\cal Z}}{\cal F}_z\eqno (1.14)$$
which makes ${\cal F}$ a ${\cal Z}$-graded algebra.
Consider the unitary operator [5]
$$E\=(-1)^{Q_g},\quad E^2=1 \eqno (1.15)$$
It induces the canonical involution $\omega$ in ${\cal F}$ by
$$\omega F\=EFE,\quad \omega^2=1 \eqno (1.16)$$
We define the bosonic part $F_b$ and the fermionic part $F_f$ of an
operator $F$ by
$$F_{b(f)}\={1\over 2}(1\,\pM\,\omega)F,\quad\Rightarrow
\omega F_{b(f)}=\,\pM\, F_{b(f)} \eqno (1.17)$$
and the graded bracket of two operators $F$ and $G$ by
$$ [F,G]=[F_b+F_f,G_b+G_f]\=[F_b,G_b]_-+[F_b,G_f]_-+[F_f,G_b]_-
+\{F_f,G_f\}_+ \eqno(1.18)$$
We also define on ${\cal F}$ the operator $d_Q$ by
$$d_QF\=[Q,F]=QF-(\omega F)Q \eqno (1.19)$$
This is a differential operator:
$$d_Q^2=0,\iff\{Q,[Q,F_b]_-\}_+=[Q,\{Q,F_f\}_+]_-=0
\eqno(1.20)$$
and an antiderivation with respect to $\omega$:
$$d_Q(FG)=(d_QF)G+(\omega F)(d_QG) \eqno (1.21)$$
The commutator relation
$$[Q_g,Q]=-Q \eqno (1.22)$$
implies
$$[\delta_g,d_Q]_-=-d_Q,\quad\Rightarrow d_Q{\cal F}_z\subseteq {\cal F}_{z-1}
\eqno (1.23)$$
i.e. $d_Q$ is a homogeneous homomorphism of degree $(-1)$ over
${\cal F}$. This implies in particular that it anticommutes with the
canonical involution:
$$\{d_Q,\omega\}_+=0 \eqno (1.24)$$
We conclude that the quadruplet $\{{\cal F},\delta_g,\omega,d_Q\}$
fits well into the definition of a graded differential algebra [6].

Let us study the action of $d_Q$ on ${\cal F}$ more explicitly. We
find
$$ d_Q\Ama(x)=i\dm\ua(x) \eqno (1.25)$$
$$d_Q\ua(x)=0,\quad d_Q\tua(x)=-i\dM\Ama(x) \eqno(1.26)$$
Eqs. (1.7, 1.8) immediately give two gauge invariants:
$$d_Q \dM\Ama(x)=d_Q\Fmna(x)=0 \eqno(1.27)$$

The above actions of $d_Q$ on {\cal F} may be called {\sl free} or
asymptotic BRS variations since the (formally defined) full
BRS variations of interacting fields [7] reduce to them in the
absence of interaction. It is exactly these free varations we are
intersted in when applying causal perturbation theory, since there
we are looking for symmetries of the $S$-matrix which is defined in
the Hilbert - Fock space $H$ of free asymptotic fields. The algebra
and homology of free BRS operators is well studied in [7,8]. The
variations induced by $d_Q$ are also called operator valued
gauge transformations in [1] since they emerge from the usual
asymptotic gauge variations in QED [12] by replacing the
gauge function $\chi$ with the ghost operator $u$. We will often
simply call this asymptotic BRS variations gauge variations
and their invariants gauge invariants.

The interaction (1.2) is gauge invariant, i.e. we
have
$$d_QT^{(1)}(x)=\dM T^{(1)\mu}(x),\quad
T^{(1)\mu}(x)\= ef_{abc}:\ua\{\ANa\Fmnc+\zw\ub\dm\tuc\}:(x) \eqno
(1.28)$$
The quintessence of causal perturbation theory is that all higher terms
$T^{(n)},\,n\geq 2$ in (1.1) are determined from $T^{(1)}$ by
Poincar\'e invariance and causality [9-12]. This determination is
unique up to some (finite!) normalization constants, which can be
determined by the requirement of symmetries and (finitely many)
normalization conditions. $T^{(n)}$ is given symbolically by
$$T^{(n)}(x_1,\cdots,x_n)=\Theta[T^{(1)}(x_1)\cdots T^{(n)}(x_n)]
\eqno(1.29)$$
where $\Theta$ means the time ordered product. This, however,
{\sl cannot} be constructed by multiplying with step functions,
since this would lead to the well known UV-divergences[9,10].
Instead one has to use the method of distribution splitting,
developed by Epstein and Glaser [11] and applied to QED, for example,
by Scharf [12]. Using exactly this construction Duetsch et al. [1-4] have shown
that
the Yang-Mills theory specified by (1.2) is gauge invariant in all
orders, i.e. the following equations hold true:
$$d_QT^{(n)}(x_1,\cdots,x_n)]=\sum_{l=1}^n\d_{\mu_l}
T^{(n)\mu_l}_{\hphantom{(n)}l}(x_1,\cdots,x_n),\eqno (1.30)$$
$$T^{(n)\mu_r}_{\hphantom{(n)}r}(x_1,\cdots,x_n)\=
\Theta[T^{(1)}(x_1)\cdots T^{(1)\mu_r}(x_r)\cdots T^{(1)}(x_n)]
\eqno(1.31)$$
Thus the gauge variation of the $T^{(n)}$ are total divergences.
One would like to conclude from this the more convential form of
gauge invariance:
$$\lim_{g\rightarrow 1}d_QS[g]=0 \eqno (1.32)$$
While this adabatic limit is well controlled in massive theories
[13] the situation is far more difficult in massless theories, where
it generally fails to exist in the $S$-matrix elements [12,14].
The strength of eq. (1.30) is to give a formulation of gauge
invariance  which is completely independent of the infrared problems
encountered when passing to the adiabatic limit.

The importance of the gauge invariance (1.30) lies in the fact that
it enables one to proof the unitary of the physical $S$-matrix $S_{\rm
phys}$ defined in the physical subspace $H_{\rm phys}$ of the total
Hilbert-Fock $H$. The latter one can be defined as the cohomology
space of $Q$ or, equivalently, as $\{{\rm Ker}Q\}\ominus_{\perp}
\{{\rm Ran}Q\}$.

The paper at hand aims at the construction of Yang-Mills theories
with massive gauge (and ghost) fields. This is usally done via the
Higgs mechanism which is known to give a renormalizable, gauge
invariant, and unitary perturbation series. While not questioning
the validity of this result we here want to develop a different
approach.

To clearly enlighten this difference let us briefly
summarize the logical steps used to derive the perturbation series
for the Higgs model. There the starting point is a classical Yang-Mills
theory defined by an action $\Sigma$ which is invariant under the
group of local classical gauge transformations $G_{\rm local}$. The
gauge fields are coupled to scalar fields which interact among each
other by a mexican hat potential. Then one considers the classical
energy which is a functional on the configuration space of classical
fields. Due to the peculiar form of the classical
potential the classical ground state, i.e. the points in the
classical configuration space which minimize the classical energy
(often very misleadingly called the vacuum), is found to be
degenerate. Then an arbitrary representative point from this state is
chosen. This is called spontaneous symmetry breaking.
After that new classical fields are defined as the original
fields shifted by this reference point.
Then a gauge fixing term is added to the action and also the
corresponding Fadeev-Popov ghost term included. The total action is
then shown to be BRS-invariant. Then the action is split
into two parts:  The free part being at most quadratically in the shifted
fields and the interaction part being at least trilinear in these
fields. It is only then that quantization comes into play:
The quadratic part of the action
defines the quantum kinematical setup, i.e. the shifted classical
fields are quantized as free quantum fields in a Hilbert-Fock space $H$
with unique (!) vacuum. These free quantized fields describe the
in- and outgoing particles. The interacting part of the
action describes the interaction of these particles and allows for
the perturbative calculation of Green functions of the corresponding
(only formally defined) interaction fields and,
most important, determines the $S$-matrix in $H$.
Again a physical subspace $H_{\rm phys}\subset H$, a formally defined
set of physical interacting observables,
and the physical $S$-matrix $S_{\rm phys}$
are defined via the homology of the BRS-transformation.
It is then shown that these physical quantities depend neither on
the representative point of the ground state of the classical energy
chosen above nor on the the gauge fixing term. Eventually $S_{\rm phys}$
is shown to be unitary. In thhis step the BRS-invariance is again the
key ingredient.
The BRS-charge is often expressed in terms of the
only formally defined interacting quantum fields. Due to its
conservation it can, however, be expressed in terms of the free
asymptotic fields as well. Since these are perfectly well defined
the latter method is superior. Moreover, it is exactly this
asymptotic BRS-invariance which is needed to proof unitarity of
$S_{\rm phys}$.

This suggests our approach to massive Yang-Mills theories: We will
not take any reference to the classical theory. So, neither the
classical gauge group $G_{\rm local}$ nor the concept of spontaneous
symmetry breaking will enter our reasoning. Instead, we
immediately start with the quantum theory , defined by given
asymptotic massive gauge and ghost fields and by the generator of the
causal $S$-matrix, $T^{(1)}(x)$. This we demand to be invariant
under asypmtotic BRS-transformations, and we give a classification
of all all $T^{(1)}(x)$  with this property. Since we do not employ
the notion of spontaneous symmetry breaking we have no reason to
include scalar fields in our discussion from the very beginning.
Instead, we derive the presence of these fields by a purely
algebraic condition.

The paper is organized as follows: The next chapter deals once more
with massless Yang-Mills theories showing that the interaction
(1.2) admits gauge invariant generalizations and giving a complete
list of them. Chapter 3 starts the investigation of massive
Yang-Mills theories. We study theories with only gauge and ghost
fields and construct their gauge invariant interactions. We will see,
however, that the BRS-charge fails to be a differential operator in
this case. Chapter 4 shows how to cure this defect: Scalar fields
have to be included in the definition of the BRS-charge to restore
its nilpotency. This changes the gauge transformations and we have
to determine the gauge invariant interactions once again. It turns
out that the scalar fields couple to the gauge fields with predicted
strength. Chapter 5 shows how to incorporate matter fields into the
theory. It is proven in chapter 6 that gauge invariance implies unitarity
of $S_{\rm phys}$. There the relation between anomalies and unitarity
is clarified, too. The critical discussion can be found in the last
chapter.

\vskip 1 cm

{\bigtype \bf 2. Gauge Invariant Interactions of Massless Yang-Mills Fields}

\vskip 0.2 cm

Here we will generalize the interaction of massless
Yang-Mills and ghost fields given by (1.1). So we have to classify their
possible interactions $T:=T^{(1)}(x)$. We will restrain the form of
these interactions by requiring it to share the following structural
properties with the interaction specified in (1.1):

\vskip 0.2cm

\ni 1.) We demand the interaction  to be normalizable. Then only normal
products of three or four fields can appear in $T$. In the case of three fields
one of them may be differentiated once.

\ni 2.) The interaction of the ghost and gauge fields shall
be of Yang-Mills type, i.e. the coupling of the gauge and ghost
fields, which are in the adjoint representation of the global Group
$G$, shall be proportional to the structure constants $f_{abc}$.

\ni 3.) We constrain the interaction $T$ to be invariant under the global
group $G$. Thus the tensor of the stucure constants has to be contracted
with three coloured fields. This excludes a posteriori the coupling of
four fields in $T$. This is quite satisfactory. For, in the next order
$T^{(2)}$, causality and gauge invariance will {\sl create} these quartic
couplings without further ado. The seagull graph in scalar QED and the four
gluon coupling in Yang-Mills theory, for example, are generated this way.

\ni 4.) We require $T$ to have vanishing ghost charge, i.e $\delta_{Q_g}T=0$.
This makes the two ghost fields $u$ and $\tu$ always appear together and
particularly implies that $T$ is a bosonic operator.

\ni 5.) We will not give up invariance of $T$ under the proper Poincar\'e
group $P_+^{\uparrow}$, of course. Thus all Lorentz indices have to be
contracted. This implies that one of the three fields coupled in $T$
has to be differentiated, because otherwise the number of Lorentz indices
would be odd.

\ni 6.) In order to have a pseudo-unitary $S$-matrix $T$ should be
anti-pseudo-hermitian.

\vskip 0.2cm

\ni There exist exactly four linear independent interaction terms $T_i$
fulfilling these conditions:
$$T_1\=: -\zw ie\f\AMa\ANb\Fmnc:\qa,\qa T_2\= :-ie\f\AMa\ub\dm\tuc:\,\,,$$
$$T_3\=-ie\f\AMa\dm\ub\tuc:\qa,\qa T_4\= :ie\f\dM\Ama\ub\tuc: \eqno(2.1)$$
and any real linear combination of these terms fulfils these conditions, too.
We therefore set
$$T=\sum_{i=1}^4 \al_iT_i \eqno(2.2)$$
with a priori arbitrary real constants $\al_i$.

Now we demand, in addition, gauge invariance, i.e.
$$d_QT=\dM\Tm \eqno(2.3)$$
Since $T$ is different from the expression (1.2), $\Tm:=T^{(1)\mu}(x)$ will be
different from (1.28), too. It shall, however, retain the following structural
properties:

\vskip 0.2cm

\ni 1'.) Normalizability: Only normal products of three or for fields may
appear in $\Tm$. In the case of three fields one may be differentiated once.

\ni 2'.) $\Tm$ shall be of Yang-Mills type, i.e. the coupling of the gauge
and ghost fields shall be proportional to the structure constants $\f$.

\ni 3'.) $G$-invarinance: All colour indices in $\Tm$ have to be contracted.
This, again, excludes the appearence of normal products of four fields in
$\Tm$.

\ni 4'.) $\Tm$ must have ghost charge $-1$. This implies that either one $u$
and no $\tu$ or two $u$ and one $\tu$ are present in $\Tm$ and that $\Tm$ is
fermionic.

\ni 5'.) $P_+^{\uparrow}$-covariance: All Lorentz indices except $\mu$ have to
be contracted.

\ni 6'.) $\Tm$ should be pseudo-hermitean.

\vskip 0.2cm

\ni These properties follow naturally from the corresponding properties 1.)-6.)
of $T$ and eq.(2.3). There exist exactly six linear independent terms
fulfilling these conditions:
$$\Tm_1\=:e\f\ua\ANb\Fmnc: \qa,\qa \Tm_2\=:-e\f\ua\Amb\dN\Anc:\,\,,$$
$$\Tm_3\=:-e\f\ua\ANb\Gmnc: \qa,\qa \Tm_4\=:\zw e\f\ua\ub\dm\tuc:\,\,,$$
$$\Tm_5\==:e\f\ua\dm\ub\tuc: \qa,\qa \Tm_6\=:e\f\dn\ua\ANb\Amc:\,\,,
\eqno(2.4)$$
Here we have introduced $\Gmna:=\dm\Ana+\dn\Ama$. Any real linear combination
of the six expressions above fulfils the requirements 1'-6', too. We therefore
set
$$\Tm=\sum_{j=1}^6\be_j\Tm_j \eqno(2.5)$$
with a priori arbitrary real constants $\be_j$.

To study (2.3) we need
$$d_QT_1=:e\f\AMa\dN\ub\Fmnc:\qa,\qa d_QT_2=:e\f\left(\AMa\ub\dm\dN\Anc+
\dM\ua\ub\dm\tuc\right):\,\,,$$
$$d_QT_3=:e\f\left(\AMa\dm\ub\dN\Anc+\dM\ua\dm\ub\tuc\right):\qa,\qa
d_QT_4=0 \eqno (2.6)$$

$$\dM\Tm_1=:e\f\left(\AMa\dN\ub\Fmnc+\AMa\ub\dm\dN\Anc\right):\qa,\qa
\dM\Tm_2=:e\f\left(\Ama\ub\dM\dN\Anc+\Ama\dM\ub\dN\Anc\right):\,\,,$$
$$\dM\Tm_3=:e\f\left(\AMa\dN\ub\Gmnc+\AMa\ub\dm\dN\Anc\right):\qa,\qa
\dM\Tm_4=:e\f\dM\ua\ub\dm\tuc:\,\,,$$
$$\dM\Tm_5=:e\f\left(\dm\ua\dM\ub\tuc+\dm\ua\ub\dM\tuc\right):\qa,\qa
\dM\Tm_6=:e\f\left(-\AMa\dm\ub\dN\Anc+\AMa\dN\ub\dm\Anc\right):\eqno(2.7)$$
To derive these formulae we have used that the fields obey the wave equation.
By inserting them in (2.3) we get a system of linear homogeneous equations for
the coefficients $\al_i$ and $\be_j$. Due to its homogeneity we can certainly
choose freely the overall normalization of these coefficients, and we do that
in setting $\al_1=1$. The solution of the equations turns out to have three
additional free parameters which we call $\al$, $\be$, and $\ga$. The general
solution is:
$$T=T_1+\left(\zw-\al\right)T_2+\left(-\zw-\al\right)T_3+\left(\al+\be\right)
T_4 \eqno(2.8)$$
$$\Tm=\left(1+\ga\right)\Tm_1+\left(-\zw-\al-2\ga\right)\Tm_2+\ga\Tm_3+\Tm_4+
\left(-\zw-\al\right)\Tm_5-2\ga\Tm_6 \eqno(2.9)$$

Let us now study the structure of these expressions. We first remark that the
special choice $\al=-\be=-\zw$ leads us back to the original interaction (1.2).
Setting, in addition, $\ga=0$ also reproduces (1.28). This choice is
distinguished by its minimality: Firstly, there are only two terms in $T$ and
$\Tm$. Secondly, only four elemantary fields are used: $A$, $F$, $u$, and
$\d\tu$. The other four elementary fields $G$, $\d\cdot A$, $\d u$, and $\tu$
do not appear at all. This shortens lengthy higher order calculations and the
very elaborate proof of gauge invariance in all orders [1-4] by a considerable
amount.

Another preferred choice is $\al=\be=\ga=0$. This gives
$$T=ie\f:\left(-\zw\AMa\Anb\Fmnc-\zw\AMa\ub\dsm\tuc\right): \eqno(2.10)$$
$$\Tm=e\f:\ua\left(\ANb\Fmnc+\zw\Amb\dN\Anc+\zw\ub\dsm\tuc\right):
\eqno (2.11)$$
This $T$ has an additional symmetry: It is invariant under the ghost charge
conjugation $C_g$ [5]. This unitary operator reflects the gauge charge:
$$C_gQ_gC_g^{-1}=-Q_g \eqno(2.12)$$
and acts on the ghost fields in the following way:
$$C_g\ua(x)C_g^{-1}=i\tua(x)\qa,\qa C_g\tua(x)C_g^{-1}=i\ua(x) \eqno (2.13)$$
This implies indeed:
$$C_gTC_g^{-1}=T \eqno(2.14)$$
and the choice $\al=\be=0$ is the only one making this equation hold true.
This $T$ is actually not only invariant under ghost charge conjugation; it is
invariant under ${\rm SU(1,1)}$ - ``rotations'' in ghost space, too [5].

The three parameter freedom in $T$, $\Tm$ has the following interpretation:

\vskip 0.2cm

\ni I.) The terms in $T$ which are multiplied by $\al$: $T_\al$, are a
pure divergence:
$$T_\al=-T_2-T_3+T_4=\dM\left(ie\f:\Ama\ub\tuc\right):\=\dM H^\mu
\eqno(2.15)$$
Since $Q$ is $x$ - independent, their gauge variation is a pure divergence,
too:
$$\d_QT_\al=d_Q\dM H^\mu=\dM\left(d_QH^\mu\right)=\dM\Tm_\al \eqno (2.16)$$
where
$$\Tm_\al\=-\Tm_2-\Tm_5 \eqno(2.17)$$
are exactly that terms in $\Tm$ which are multiplied by $\al$. It follows that
the couple $(T_\al,\Tm_\al)$ fulfils (2.3) separately.

\ni II.) The term in $T$ which is  multiplied by $\be$: $T_\be$, is a pure
gauge, i.e. a $d_Q$-boundary:
$$T_\be=T_4=d_QL\qa,\qa L\=-\zw e\f:\ua\tub\tuc: \eqno (2.18)$$
Since $d_Q$ is a differential operator it is also a $d_Q$-cycle, i.e gauge
invariant:
$$d_QT_\be=0 \eqno(2.19)$$
It can, therefore, freely be added to $T$ without invalidating (2.3).

\ni III.) The terms in $\Tm$ which are multiplied by $\ga$: $\Tm_\ga$, are a
conserved trilinear current:
$$\Tm_\ga=\Tm_1-2\Tm_2+\Tm_3-2\Tm_6=2\dN\{:e\f\ua\Amb\Anc:\}
\=K^\mu\qa,\qa \dM K^\mu=0 \eqno(2.20)$$
It follows that they can freely be added to $\Tm$ without invalidating (2.3).

\vskip 0.2cm

The discussion above allows to reformulate (2.8, 2.9) as
$$T=T_1+\zw\left(T_2-T_3\right)+\al\,\dM H^\mu+\be\,d_QL\,\,,$$
$$\Tm=\Tm_1+\Tm_4-\zw\left(\Tm_2+\Tm_5\right)+\al\,d_QH^\mu+\ga\,K^\mu
\eqno(2.21)$$
In the next chapter we will study how these structures change if the gauge
fields and ghost fields are massive.

\vfill\eject

{\bigtype \bf 3. A Direct Route to Massive Yang-Mills Fields}

To construct a theory of massive Yang-Mills fields we have to use free
asymptotic massive gauge and ghost fields:

$$(\dq+M^2)\Ama(x)=(\dq+M^2)\ua(x)=(\dq+M^2)\tua(x)=0 \eqno (3.1)$$
$$[\Ama(x),\Anb(y)]_-=i\delta_{ab}g^{\mu\nu}D_M(x-y) \eqno (3.2)$$
$$\{\ua(x),\tub(y)\}_+=-i\delta_{ab}D_M(x-y),\,\,\,
\{\ua(x),\ub(y)\}_+=\{\tua(x),\tub(y)\}_+=0 \eqno (3.3)$$
where  $D_M$, the Pauli-Jordan commutation function for mass $M>0$, appears.
The free massive field strenghts $\Fmna$ are defined as in (1.4).
We have given all coloured fields the same mass, since we do not discuss
breaking of the global group $G$ here, while the ghost  and the gauge fields
have the same mass because they transform among each other under gauge
transformations.

The nonvanishing of the mass $M$ has simple but far reaching consequences:
While (1.8) remains true,
$$[\dM\Ama(x),F_b^{\kappa\lambda}(y)]_-=0 \eqno(3.4)$$
(1.7) is altered to
$$[\dM\Ama(x),\dN\Anb(y)]_-=iM^2\delta_{ab}D_M(x-y) \eqno(3.5)$$
Let us define the gauge charge $Q$ by
$$Q\=\intx\left(\dM\Ama(x)\right)\dsO \ua(x) \eqno(3.6)$$
While this is the same {\sl expression} as in the massless case, it is, of
course, a different operator, because now the quantized fields in the integral
are the massive ones. Its square is given by
$$Q^2=\zw\{Q,Q\}_+={1\over 2}\intx\inty\cdot$$
$$\cdot\Bigl\{[\dM\Ama(x),\dN\Anb(y)]_-\dsOx\dsOy\left(\ua(x)\ub(y)\right)+
\left(\dN\Anb(y)\dM\Ama(x)\right)\dsOx\dsOy\{\ua(x),\ub(y)\}_+\Bigr\}
\eqno(3.7)$$
and this is due to the nonvanishing commutator (3.5) unequal to zero,
in contrast to the massless case (1.10). Instead it is given by
$$Q^2=iM^2Q_u\qa,\qa Q_u\=i\intx \ua(x)\dso\ua(x) \eqno(3.8)$$
The charge $Q_u$ has been discussed in the framework of the ghost charge
algebra in [5]. So $Q$ fails to be a differental operator and homological
notions do not apply. This applies as well to the gauge variation $d_Q$, which
is defined by (1.19) also in the massive case: (1.20) is changed to
$$d_Q^2=iM^2\delta_u \eqno(3.9)$$
where $\delta_u$ is the derivation induced by $Q_u$. The algebraic Eqs.
(1.11-1.19, 1.21-1.24) remain true in the massive theory, while the gauge
variation of the basic fields changes from (1.25)-(1.27) to
$$d_Q\Ama(x)=i\dm\ua(x)\,\,,\,\,d_Q\Fmna(x)=0\,\,,\,\,
d_Q\left(\dM\Ama(x)\right)=iM^2\ua(x)\,\,, \eqno(3.10)$$
$$d_Q\ua(x)=0\,\,,\,\,d_Q\tua(x)=-i\dM\Ama(x) \eqno(3.11)$$

Now we look again for the general gauge invaraint interaction, i.e. any couple
$(T,\Tm)$ fulfilling the general conditions discussed in the preceding chapter
and $d_QT=\dM\Tm$. We can take over the expressions (2.1) and (2.4) and the
Ans\"atze (2.2) and (2.5). Since, however, the fields in these expressions are
now massive (2.6) and (2.7) change to
$$d_QT_1=:e\f\AMa\dN\ub\Fmnc:\qa,\qa d_QT_2=:e\f\left(\AMa\ub\dm\dN\Anc+
\dM\ua\ub\dm\tuc\right):\,\,,$$
$$d_QT_3=:e\f\left(\AMa\dm\ub\dN\Anc+\dM\ua\dm\ub\tuc\right):\qa,\qa
d_QT_4=:e\f M^2\ua\ub\tuc: \eqno (3.12)$$
$$\dM\Tm_1=:e\f\left(\AMa\dN\ub\Fmnc+\AMa\ub\dm\dN\Anc\right):\qa,\qa
\dM\Tm_2=:e\f\left(\Ama\ub\dM\dN\Anc+\Ama\dM\ub\dN\Anc\right):\,\,,$$
$$\dM\Tm_3=:e\f\left(\AMa\dN\ub\Gmnc+\AMa\ub\dm\dN\Anc\right):\qa,\qa
\dM\Tm_4=:e\f\left(\dM\ua\ub\dm\tuc-\zw M^2\ua\ub\tuc\right):\,\,,$$
$$\dM\Tm_5=:e\f\left(\dm\ua\dM\ub\tuc+\dm\ua\ub\dM\tuc-M^2\ua\ub\tuc\right):
\,\,,$$
$$\dM\Tm_6=:e\f\left(-\AMa\dm\ub\dN\Anc+\AMa\dN\ub\dm\Anc\right):\eqno(3.13)$$
This leads to the result
$$T=T_1+\zw(T_2-T_3)+\al\,\dM H^\mu,$$
$$\Tm=\Tm_1+\Tm_4-\zw\left(\Tm_2+\Tm_5\right)+\al\,d_QH^\mu+\ga\,K^\mu
\eqno(3.14)$$
where $H^\mu$ and $K^\mu$ are the expressions defined in (2.15) and (2.20).

The result has the same structure and interpretation as (2.21) in the massless
theory, but for one difference: There is no free term of the form $\be\,d_QL$
here. The reason is clear: Since $d_Q$ fails to be a differential operator
such a term would not be gauge invariant. Since this term is absent, no
analogon of the minimal choice exists in the massive theory. A $C_g$ -
symmetric interaction $T$, however, does exist, and is again uniquely given by
setting $\al=0$. For this case, the expressions $T$ and $\Tm$ are identical to
those in the massless case. So one might say that this additional symmetry has
stabilized the theory against perturbations by mass terms.

We have succeeded in constructing gauge invariant interactions for massive
Yang-Mills fields. This theory, however, has no physical interpretation.
For, we will see in chapter six that the not abundanable unitarity of the
physical $S$-matrix is only guaranteed if both equations: $d_QT=\dM\Tm$ and
$Q^2=0$ hold true, but the latter one fails here. These problems have also been
studied in the canonical framework [7].

We would like to point out that the theory defined above may none the
less be quite useful: It is a gauge invariant infrared regulator for the
massless theory, and a properly done comparsion of the two can give important
results.

Knowing now exactly where the straightforward approach to massive quantized
Yang - Mills fields fails we will not find it to difficult to cure this
problem. This is done in the next chapter.

\vskip 1cm

{\bigtype\bf 4. The Algebraic Introduction of Scalar Gauge Fields}

\vskip 0.2cm

{}From (3.7) we learn that the the reason why $Q$ fails to be a differential
operator is simply the nonvanishing commutator (3.5). Consider now
${\rm dim}\,G$ hermitean free quantized Klein-Gordon fields $\ha(x)$ which
are, like the gauge fields $\Ama$ and the ghosts fields $\ua$ and $\tua$,
in the adjoint representation of $G$ and which have the same mass as these
fields. They obey
$$(\dq+M^2)\ha(x)=0\,\,,\,\,[\ha(x),\hb(y)]_=-i\dab D_M(x-y) \eqno(4.1)$$
Their commutuator has the opposite sign to (3.5). It follows that the fields
$\dM\Ama(x)+M\ha(x)$ have vanishing commutators with themselves:
$$[\dM\Ama(x)+M\ha(x),\dN\Anb(y)+M\hb(y)]_-=0 \eqno (4.2)$$
This suggests the following new definition for $Q$:
$$Q\=\intx\left(\dM\Ama(x)+M\ha(x)\right)\dsO\ua(x) \eqno(4.3)$$
For, this implies
$$Q^2={1\over 2}\intx\inty\Bigl\{[\dM\Ama(x)+M\ha(x),\dN\Anb(y)+M\hb(y)]_-
\dsOx\dsOy\left(\ua(x)\ub(y)\right)+$$
$$+\bigl(\dN\Anb(y)+M\hb(y)\bigr)\bigl(\dM\Ama(x)+M\ha(x)\bigr)
\dsOx\dsOy\{\ua(x),\ub(y)\}_+\Bigr\}=0 \eqno(4.4)$$
So we have managed to recover this important algebraic property which, together
with gauge invariance, guarantees unitarity of $S_{\rm phys}$! Note that all
eqs. (1.11-1.20) hold true anew.

The gauge variations of the elementary fields are given by
$$d_Q\Ama(x)=i\dm\ua(x)\,\,,\,\,d_Q\ha(x)=iM\ua(x)\,\,,\,\,d_Q\ua(x)=0\,\,,\,\,
d_Q\tua(x)=-i\bigl(\dM\Ama(x)+M\ha(x)\bigr) \eqno(4.5)$$
Consequently,
$$ d_Q\dM\Ama(x)=-iM^2\ua(x)\,\,,\,\, d_Q\Fmna(x)=d_Q\bigl(\dM\Ama(x)+M\ha(x)
\bigr)=0 \eqno(4.6)$$
We see that the scalar fields $\ha(x)$ are effected by the gauge variation
$d_Q$ and that they appear in the gauge variations of other fields. Hence
it is appropriate to call them scalar gauge fields. We will see in chapter six
that these fields are unphysical, i.e. their projections onto $H_{\rm phys}$
vanish.

We now have to determine the possible gauge invariant interactions $(T,\Tm)$
once more. We will not dispense with the structural conditions discussed in
chapter 2. These conditions, however, can now be fulfiled by more expressions
$T_i$ and $\Tm_i$ than in the preceding chapters, since the presence of the
scalar fields allows for the constructions of new terms. We give the following
complete list:
$$T_1\=: -\zw ie\f\AMa\ANb\Fmnc:\,\,,\,\,
  T_2\= :-ie\f\AMa\ub\dm\tuc:\,\,,\,\,
  T_3\=-ie\f\AMa\dm\ub\tuc:\,\,,$$
$$T_4\= :ie\f\dM\Ama\ub\tuc:\,\,,\,\,
  T_5\=:ie\f M\ha\ub\tuc:\,\,,\,\,
  T_6\=:\zw ie\f\AMa\hb\dsm\hc: \eqno(4.7)$$
$$\Tm_1\=:e\f\ua\ANb\Fmnc:\,\,,\,\,
  \Tm_2\=:-e\f\ua\Amb\dN\Anc:\,\,,\,\,
  \Tm_3\=:-e\f\ua\ANb\Gmnc:\,\,,$$
$$\Tm_4\=:\zw e\f\ua\ub\dm\tuc:\,\,,\,\,
  \Tm_5\==:e\f\ua\dm\ub\tuc:\,\,,\,\,
  \Tm_6\=:e\f\dn\ua\ANb\Amc:\,\,,$$
$$\Tm_7\=:-e\f M\ua\Amb\hc:\,\,,\,\,
  \Tm_8\=-\zw e\f\ua\hb\dsm\hc \eqno(4.8)$$
Using
$$d_QT_1=:e\f\AMa\dN\ub\Fmnc:\,\,,\,\,
  d_QT_2=:e\f\left(\AMa\ub\dm\bigl[\dN\Anc+M\hc\bigr]+\dM\ua\ub\dm\tuc\right):
$$
$$d_QT_3=:e\f\left(\AMa\dm\ub\bigl[\dN\Anc+M\hc\bigr]+\dM\ua\dm\ub\tuc\right):
\,\,,\,\,
  d_QT_4=:e\f\left(-M\dM\Ama\ub\hc+M^2\ua\ub\tuc\right):\,\,,$$
$$d_QT_5=:e\f\left( M\dM\Ama\ub\hc-M^2\ua\ub\tuc\right):\,\,,$$
$$d_QT_6=:e\f\left(-\dM\ua\hb\dm\hc-M\AMa\ub\dm\hc-M\AMa\hb\dm\uc\right):
\eqno (4.9)$$
$$\dM\Tm_1=:e\f\left(\AMa\dN\ub\Fmnc+\AMa\ub\dm\dN\Anc\right):\,\,,\,\,
  \dM\Tm_2=:e\f\left(\Ama\ub\dM\dN\Anc+\Ama\dM\ub\dN\Anc\right):\,\,,$$
$$\dM\Tm_3=:e\f\left(\AMa\dN\ub\Gmnc+\AMa\ub\dm\dN\Anc\right):\,\,,\,\,
  \dM\Tm_4=:e\f\left(\dM\ua\ub\dm\tuc-\zw M^2\ua\ub\tuc\right):\,\,,$$
$$\dM\Tm_5=:e\f\left(\dm\ua\dM\ub\tuc+\dm\ua\ub\dM\tuc-M^2\ua\ub\tuc\right):
\,\,,$$
$$\dM\Tm_6=:e\f\left(-\AMa\dm\ub\dN\Anc+\AMa\dN\ub\dm\Anc\right):\,\,,$$
$$\dM\Tm_7=:e\f M\left(\dM\Ama\ub\hc+\Ama\dM\ub\hc+\Ama\ub\dM\hc\right):\,\,,$$
$$\dM\Tm_8=:e\f\ha\dM\ub\dm\hc \eqno(4.10)$$
we find the general gauge invariant interaction as follows:
$$T=T_1+(\zw-\al)T_2-(\zw+\al)T_3+(\al+\be)T_4+\be\,T_5+\zw T_6=$$
$$= T_1+\zw(T_2-T_3+T_6)+\al\,\dM H^{\mu}+\be\,d_QL \eqno(4.11)$$
$$\Tm=(1+\ga)\Tm_1-(\zw+\al+2\ga)\Tm_2+\ga\,\Tm_3+\Tm_4-(\zw+\al)\Tm_5
     -2\ga\,\Tm_6-\al\,\Tm_7+\zw\Tm_8=$$
$$   =\Tm_1+\Tm_4+\zw(\Tm_8-\Tm_2-\Tm_5)+\al\,d_QH^{\mu}+\ga\,K^{\mu}
\eqno(4.12)$$
where
$$H^{\mu}\=:ie\f\Ama\ub\tuc:\,\,,\,\,L\=:-\zw e\f\ua\ub\tuc:\,\,,\,\,
  K^{\mu}\= 2\dN\left\{:e\f\ua\Amb\Anc:\right\} \eqno(4.13)$$

We notice that the term $T_6$ enters in $T$ with a fixed coefficient. This
term describes the interaction of the scalar gauge fields $\ha$ with the
Yang-Mills fields $\Ama$. Thus the strength of this interaction is fixed by
the condition of gauge invariance. The free terms multiplied by $\al$, $\be$,
and $\ga$ have the same interpretation as in the massless case: $\dM H^{\mu}$
is a pure divergence, $d_QL$ is pure gauge, and $K^{\mu}$ is a conserved
current. The condition of $C_g$-invariance on $T$ uniquely fixes $\al=\be=0$.
Since the current $K^{\mu}$ is not related to the gauge structure of the
theory, $\ga=0$ is certainly a sensible choice, too. In this case the
interaction is given explicitly by
$$T=ie\f:\left\{-\zw\AMa\ANb\Fmnc-\zw\AMa\ub\dsm\tuc+{1\over 4}\AMa\hb\dsm\hc
\right\}\,\,,$$
$$\Tm=e\f:\ua\left\{\ANb\Fmnc+\zw\ub\dsm\tuc+\zw\Amb\dN\Anc-{1\over 4}\hb\dm\hc
\right\}: \eqno(4.14)$$

We have now succeeded in constructing gauge invariant interactions of
Yang-Mills fields, ghost fields, and scalar gauge fields. In the next chapter
we study how these fields couple to matter fields.

\vskip 1cm

{\bigtype\bf 5. The Coupling of Matter Fields}

All fields we have studied so far, the Yang-Mills fields $\Ama$, the scalars
$\ha$, and the ghosts $\ua$ and $\tua$ may be called gauge fields, since they
transformform among each other under the gauge variation $d_Q$. Now we will
study additional fields, which we call matter fields. We will study scalar
matter fields: $\Fi$ (adjoint fields: $\Ki$) and Dirac matter fields: $\Pi$
(Dirac-adjoint fields: $\Qi=\psi^+_i\ga_5$). $i$ is an internal index
numbering different fields, while the bispinor indices $\al$ for the Dirac
fields are always supressed.  Let us assume that the fields $\Fi,\Pi$ transform
under a certain irreducible representation$R$ of $G$ in which the hermitean
generators are given by $R_a^{ij}$. The matter fields will form bilinear
currents in the following way:

1.) The scalar fields constitute the currents
$$\sma=:\Ki R_a^{ij}i\dsm\Fj: \eqno(5.1)$$
If we demand strictly $G$-invariance, all fields $\Fi$ should have the same
mass $m_S$:
$$(\dq+m_S^2)\Fi=0 \eqno(5.2)$$
In this case the currents are conserved:
$$\dM\sma=0 \eqno(5.3)$$
Let us, however allow for $G$-breaking in the mass sector of the scalar matter
fields, i.e. replace (5.2) by
$$\left(\dq\,\de^{ij}+\mas^{ij}\right)\Fj=0 \eqno(5.4)$$
where $\mas$ is the positive mass-square matrix of the scalar fields in $R$.
Then (5.3) does not hold. Instead we find
$$\dM\sma=:\Ki i\left[\mas,R_a\right]_-^{ij}\Fj: \eqno(5.5)$$

2.) The Dirac fields form two kinds of currents. The vector currents are
defined by
$$\vma=:\Qi R_a^{ij}\ga^\mu\Pj: \eqno(5.6)$$
while the axial currents are given by
$$\xma=:\Qi R_a^{ij}\ga^\mu\ga_5\Pj: \eqno(5.7)$$
Strict $G$-invariance would require that all Dirac fields in $R$ have
the same mass $m_D$:
$$(i\,\ga^\mu\dM-m_D)\Pi=0 \eqno(5.8)$$
In this case the vector current would be conserved and the divergence of the
axial current would be the pseudo-scalar
$$\dM\vma=0\,\,,\,\,\dM\xma=2m_D:\Qi R_a^{ij}i\,\ga_5\Pj: \eqno(5.9)$$
Let us, however allow for $G$-breaking in the mass sector of the Dirac matter
fields, i.e. replace (5.8) by
$$\left(i\,\ga^\mu\dM\,\de^{ij}-\mad^{ij}\right)\Pj=0 \eqno(5.10)$$
where $\mad$ is the hermitean mass matrix of the Dirac fields in $R$.
Then (5.9) does not hold. Instead we find
$$\dM\vma=:\Qi i\left[\mad,R_a\right]_-^{ij}\Pj:\,\,,\,\,
  \dM\xma=:\Qi\left\{\mad,R_a\right\}_+^{ij}i\,\ga_5\Pj: \eqno(5.11)$$

The interaction $T$, $\Tm$ constructed in the last chapter contains only gauge
fields. So let us call it $\TG$, $\TGm$ from now on. We now add to it $\TM$,
$\TMm$ to desribe the interaction of the matter fields with the ghost fields:
$$T=\TG+\TM\,\,,\,\, \Tm=\TGm+\TMm \eqno(5.12)$$
$\TM$ is constructed by coupling the above currents to the gauge fields: Let
us introduce the total current
$$\jma\=i(e_S\sma+e_V\vma+e_X\xma) \eqno(5.13)$$
and set
$$\TM=i\jma\AMa+\cdots \eqno(5.14)$$
The dots indicate that we will soon add other terms to this expression.
For, let us study gauge invariance. Since $Q$ (4.3) is entirely composed
of gauge fields it commutes with the matter fields and their currents:
$$d_Q\Fi=d_Q\Pi=d_Q\jma=0 \eqno(5.15)$$
This leads to
$$d_Q\TM=-\jma\dM\ua+d_Q(\cdots)=\dM\left\{-\jma\ua\right\}+
\left(\dM\jma\right)\ua+d_Q(\cdots) \eqno(5.16)$$
The first term on the right hand side of this equation is already a divergence.
The second term vanishes iff the currents $\jma$ are conserved but is not a
divergence otherwise. So it has to be compensated by the third term, i.e. we
get the condition
$$d_Q(\cdots)=-\left(\dM\jma\right)\ua \eqno(5.17)$$
which is easily solved by
$$\cdots=iM^{-1}\left(\dM\jma\right)\ha \eqno(5.18)$$
Such we have found the following gauge invariant interaction of the matter
fields with the gauge fields:
$$\TM=i\left\{\jma\AMa+M^{-1}\dM\jma\ha\right\}\,\,,\,\,
\TMm=-\jma\ua\,\,,\,\,d_Q\TM=\dM\TMm \eqno(5.19)$$

Let us interpret this result. In the case of conserved currents the matter
fields couple only to the Yang-Mills fields and this interaction has the same
form as the coupling of matter fields to massless Yang-Mills fields [15].
More interesting is the case of nonconserved currents: There the matter fields
couple to the scalar gauge fields, too. We conclude that the scalar gauge
fields are a very important part of the whole theory:
They allow a consistent treatment
of massive Yang-Mills fields and of nonconserved currents at the same time.
We also notice that the coupling of the nonconseved currents to the scalar
gauge fields is proportional to the inverse mass of the gauge fields. That is
only possible if this mass does not vanish, which is in agreement
with experiment: The conserved strong vector currents couple to massless
(though confined) gluons while the nonconserved weak axial currents couple
to the massive weak bosons!

Let us remark that the currents $\sma$, $\vma$, and $\xma$ couple with
different coupling constants $e_S$, $e_V$, and $e_X$ to the Yang-Mills fields,
as described in (5.13, 5.14). We also have always used the same
irreducible reprentation $R$ for the matter fields. This is, of course, not
necessary: The scalar fields will generally occur in other representations
than the Dirac fields and the vector currents will generally be made out of
fermions in representations different from those to which the fermions
constituting the axial currents belong. Moreover, the representations $R$
need not to be irreducible. Instead, they can be the direct sum of several
irreducible parts. The above formulae remain true if one defines the total
current (5.13) as the sum of all currents in the various representations $R$.
Then each representation $R$ can have its own coupling constant $e_R$.
We remark, however, that this is a typical first order phenomenon. The
condition of gauge invariance in second order will certainly give
restrictions on the various coupling constants which seem to arbitrary at the
moment. This also happens in massless Yang-Mills theory, where the gauge
invariance of certain second order tree graphs implies the equality of the
Yang-Mills self-coupling constant with the one in the coupling of the
Yang-Mills fields to matter[15].

We now have carefully studied the possible interactions of {\sl quantized}
Yang-Mills fields, ghost fields, scalar gauge fields, and scalar and Dirac
matter fields. Though only working in first order $T^{1}$ we have discovered
very interesting structures. To complete the theory, we would have to study
gauge invariance in all orders, eq.(1.30). Before we take on this Herculean
task we like to know what we get if we succeed. It is the unitarity of
$S_{\rm phys}$. This is shown in the next chapter.

\vskip 1cm

{\bigtype\bf 6. Gauge Invariance and Unitarity of the Physical $S$-Matrix}

\vskip 0.2cm

Unitarity of the physical $S$-matrix in the case of massless Yang-Mills fields
was proven in [4]. Here we treat the massive case, i.e. the interaction
constructed in the two preceding chapters.

Let us begin with discussing the Krein structure [5,8,16,17]
in the Hilbert-Fock space of the gauge fields. The massive Yang-Mills fields
are quantized as
$$\Ama(x)=\sum_{\la=0}^3\intk\left\{\ep^\mu_\la(k)a_{\la,a}(k)\emi
+\ep^\mu_\la(k)a^K_{\la,a}(k)\ei\right\} \eqno(6.1)$$
$k$ is always on the mass shell ${\cal M}$:
$$k\=(k_0,\vec{k})\,\,,\,\,k_0\=+[(\vec{k})^2+M^2]^\zw\,\,,\,\,
dk\={d^3\vec{k}\over 2k_0(2\pi)^3}\,\,,\,\, \de(k-k')\=2k_0(2\pi)^3\de^{(3)}
\left(\vec{k}-{\vec{k}}'\right) \eqno(6.2)$$
$\ep_\la^\mu$ are four polarisation vectors satisfying
$$\ep_0^\mu(k)\={k^\mu\over M}\,\,,\,\,g_{\mu\nu}\ep_\la^\mu(k)\ep_\ka^\nu(k)=
g_{\ka\la}\,\,,$$
$$\sum_{\la=1}^3\ep_\la^\mu(k)\ep_\la^\nu(k)=
-\left[g^{\mu\nu}-{k^\mu k^\nu\over M^2}\right]\,\,,\,\,
\sum_{\la=0}^3g_{\la\la}\ep_\la^\mu(k)\ep_\la^\nu(k)=g^{\mu\nu}\,\,,\,\,
\overline{\ep_\la^\mu(k)}=\ep_\la^\mu(k) \eqno(6.3)$$
$a_{\la,a}(k)$ are $4({\rm dim}\,G)$
standard (distributional) bosonic annihilation operators [12,17,18,19] acting
in the Hilbert- Fock space
$$H_A=\bigoplus_{n=0}^\infty\left\{\bigvee^n\left\{\mathop{\oplus}_{\la=0}^3
\mathop{\oplus}_{a=1}^{{\rm dim}\,G}\left[L^2({\cal M},dk)\right]_{\la,a}
\right\}\right\} \eqno(6.4)$$
which is equipped with the standard positive scalar product $(\Ua,\Ub)_A$.
The operator $O^+$ denotes the adjoint of $O$ with respect to this scalar
product. The Fock space operators fulfil
$$[a_{\la,a}(k),a^+_{\ka,b}(k')]_-=\de_{\la\ka}\de_{ab}\de(k-k') \eqno(6.5)$$
The number operators for a given polarization $\la$ are defined by
$$N_\la\=\intk a_{(\la),a}^+(k)a_{(\la),a}(k) \eqno(6.7)$$
The Krein operator $J_A$ in $H_A$ [5,8,16,17] is defined by
$$J_A\=(-1)^{N_0} \eqno(6.8)$$
It defines a pseudo-conjugation $O^K$ [4,5,12,16,17] of an operator $O$ by
$$O^K\=J_AO^+J_A \eqno(6.9)$$
Sometimes the form
$<\Ua,\Ub>_A:=(\Ua,J_A\Ub)_A$
is called an indefinite scalar product. We will not follow this terminology
here. The word orthogonal (hermitean, unitary) will always mean orthogonal
(hermitean, unitary) with respect to the positive inner product. Otherwise we
say pseudo-orthogonal (pseudo-hermitean, pseudo-unitary).

The gauge invariant {\sl physical} Yang-Mills fields $A_{\rm phys}$ have only
the three transversal polarisations:
$$\Amap(x)=\sum_{\la=1}^3\intk\left\{\ep^\mu_\la(k)a_{\la,a}(k)\ei
+\ep^\mu_\la(k)a^+_{\la,a}(k)\emi\right\}\,\,,\,\,d_QA_{\rm phys}=0
\eqno(6.10)$$
and the commutator
$$[\Amap(x),\Anbp(y)]_-=-\left[g^{\mu\nu}+{\dm\dn\over M^2}
\right]\de_{ab}(-i)D_m(x-y) \eqno(6.11)$$
The {\sl unphysical} Yang Mills $A_{\rm unphys}$ fields are given by
$$\Amau(x)\=\Ama(x)-\Amap(x)={-1\over M^2}\dm\dN\Ana(x) \eqno(6.12)$$
The following conjugation properties are easily checked:
$$A=A^K\,\,,\,\,A_{\rm phys}=A_{\rm phys}^K=A_{\rm phys}^+\,\,,\,\,
A_{\rm unphys}=A_{\rm unphys}^K=-A_{\rm unphys}^+ \eqno(6.13)$$
We also note
$$\Fmna\=\dm\Ana-\dn\Ama=\dm\Anap-\dn\Amap=(\Fmna)^K=(\Fmna)^+ \eqno(6.14)$$
$$\dM\Ama=\dM\Amau=\left(\dM\Ama\right)^K=-\left(\dM\Ama\right)^+ \eqno(6.15)$$

The representation of the proper Poincar\'e group
$P_+^\uparrow$ in $H_A$ is defined by
$$U_A(a,\Lambda)\Ama(x)U_A(a,\Lambda)^{-1}=
\Lambda^\mu_{\hphantom{\mu}\nu}\Ana(\Lambda x+a)\,\,,\,\, U_A(a,\Lambda)
\Omega_A=\Omega_A \eqno(6.16)$$
where $\Omega_A$ is the vacuum in $H_A$. It is pseudo-unitary:
$$U_A(a,\Lambda)U_A(a,\Lambda)^K=1 \eqno(6.17)$$
and, since it commutes with $J_A$:
$$J_AU_A(a,\Lambda)J_A=U_A(a,\Lambda) \eqno(6.18)$$
unitary as well:
$$U_A(a,\Lambda)U_A(a,\Lambda)^+=1 \eqno(6.19)$$
The last two equations fail in the massless case where $\ep_0^\mu(k)$ cannot
be chosen covariantly.

Next we come to the hermitean scalar gauge fields which are quantized in the
usual way:
$$\ha(x)= \intk \left\{b_a(k)\emi+b_a^+(k)\ei\right\}=\ha^+(x) \eqno(6.20)$$
as operators in the Hilbert-Fock space
$$H_h=\bigoplus_{n=0}^\infty\left\{\bigvee^n\left\{\mathop{\oplus}_{a=1}
^{{\rm dim}\,G}\left[L^2({\cal M},dk)\right]_a
\right\}\right\} \eqno(6.21)$$
with standard positive scalar product $(\Ua,\Ub)_h$. We have
$$[b_a(k),b_b^+(k')]_-=\de_{ab}\de(k-k') \eqno(6.22)$$
We do not introduce an additional Krein structure in $H_h$. This is equivalent
to saying that $J_h=1$ and that the two conjugations agree over $H_h$:
$O^K=O^+$. The same is true for the two forms over $H_h$:
$<\Ua,\Ub>_h=(\Ua,\Ub)_h$. The representation of $P_+^\uparrow$ in
$H_h$: $U_h(a,\Lambda)$ is unitary. The total number of scalar gauge particles
is given by
$$N_h=\intk b^+_a(k)b_a(k) \eqno(6.23)$$

Now we consider the ghost fields. They have been extensively studied
in [5]. So we summarize only the most important formulae here. The ghost
fields
$$\ua(x)=\intk\left\{c_{1,a}(k)\emi+c_{-1,a}^+(k)\ei\right\}\,\,,\,\,
\tua(x)=\intk\left\{-c_{-1,a}(k)\emi+c_{1,a}^+(k)\ei\right\} \eqno(6.24)$$
are defined in the Hilbert-Fock space
$$H_g=\bigoplus_{n=0}^\infty\left\{\bigwedge_n\left\{\mathop{\oplus}_{i=\pm 1}
\mathop{\oplus}_{a=1}^{{\rm dim}\,G}\left[L^2({\cal M},dk)
\right]_{i,a}\right\}\right\} \eqno(6.25)$$
with positive inner product $(\Ua,\Ub)_g$. The index $i$ distinguishes ghost
from antighost particles. We have
$$\{c_{i,a}(k),c^+_{j,b}(k')\}_+=\de_{ij}\de_{ab}\de(k-k') \eqno(6.26)$$
The Krein operator in $H_g$ is defined by
$$J_g=i^{N_g-\Gamma} \eqno(6.27)$$
Here $N_g$ denotes the total number of ghost and antighost particles:
$$N_g=\intk c_{i,a}^+(k)c_{i,a}(k) \eqno(6.28)$$
while $\Gamma$ is defined by
$$\Gamma=\intk\left\{c_1^+(k)c_{-1}(k)+c_{-1}^+(k)c_1(k)\right\} \eqno(6.29)$$
Again one considers the indefinite form $<\Ua,\Ub>_g:=(\Ua,J_g\Ub)$ and defines
$O^K=J_gO^+J_g$. The representation of $P_+^\uparrow$ in $H_g$:
$U_g(a,\Lambda)$ is unitary and,
since it commutes with $J_g$, pseudo-unitary as well [5].

The scalar and Dirac matter fields are quantized in their own Hilbert-Fock
space $H_{\rm matter}$ in the usual way. The scalar product in this space is
again positive and the Krein structure $J_{\rm matter}$ is the unit operator.
The representation $U_{\rm matter}(a,\Lambda)$ is unitary.

The Hilbert space $H$ of the total system is the tensor-product of the spaces
above:
$$H=H_A\otimes H_h\otimes H_g\otimes H_{\rm matter} \eqno(6.30)$$
The Krein operator and the representation of $P_+^\uparrow$ factorize
accordingly:
$$J=J_A\otimes J_h\otimes J_g\otimes J_{\rm matter} \eqno(6.31)$$
$$U(a,\Lambda)=U_A(a,\Lambda)\otimes U_h(a,\Lambda)\otimes U_g(a,\Lambda)
\otimes U_{\rm matter}(a,\Lambda) \eqno(6.32)$$
This $U$ is unitary and pseudo-unitary, since it commutes with $J$. The
positive scalar product in $H$ is denoted by $(\Ua,\Ub)$ and
$||\Ua||:=(\Ua,\Ua)$, $<\Ua,\Ub>:=(\Ua,J\Ub)$, $O^K:=JO^+J$.

Our next task is to study more closly the gauge charge $Q$ (4.3).
It is expressed in momentum space as
$$Q=M\intk \left\{c^+_{-1,a}(k)\left[a_{0,a}(k)+ib_a(k)\right]
-\left[a^+_{0,a}(k)+ib^+_a(k)\right]c_{1,a}(k)\right\} \eqno(6.33)$$
Its adjoint $Q^+$ is given by:
$$Q^+=M\intk\left\{c^+_{1,a}(k)\left[-a_{0,a}(k)+ib_a(k)\right]
+\left[a^+_{0,a}(k)-ib^+_a(k)\right]c_{-1,a}(k)\right\} \eqno(6.34)$$
$Q$ and $Q^+$ are both pseudo-hermitean $P_+^\uparrow$ invarinat differential
operators:
$$Q^2=\left(Q^+\right)^2=0\,\,,\,\,Q=Q^K\,\,,\,\,\left(Q^+\right)^K=Q^+
\,\,,\,\,U(a,\Lambda)Q^{(+)}U(a,\Lambda)^{-1}=Q^{(+)} \eqno(6.35)$$

We now follow Razumov and Rybkin [8] who showed that the physical Hilbert
space of a gauge theory with quadratic BRS charge $Q$ can be defined as
$$\hp\={\rm kernel}\,\{Q,Q^+\}_+ \eqno(6.36)$$
Razumov showed the equivalence of this definition  with the more conventional
one using equivalent classes in semidefinite metric spaces [7]. Razumovs
definition is advantageous because it realizes $\hp$ as a concrete
subspace of the Hilbert space $H$ which has a clear particle interpretation.
To work this out we only have to calculate the above anticommutator. We find:
$$\{Q,Q^+\}_+=2[N_0+N_h+N_g]\=2N \eqno(6.37)$$
i.e. $N$ is the number of longitudinal Yang-Mills fields plus the number of
scalar gauge fields plus the number of ghost and antighost particles. Thus all
these particles are unphysical. The only physical particles are the transverse
quanta of the Yang-Mills fields and the matter particles.

The spectrum of the number operator $N$ are the natural numbers and $0$:
$$N=\sum_{n=0}^{\infty}nP_n \eqno(6.38)$$
where $P_n$ is the orthogonal projector on the subspace with $n$ unphysical
particles. (6.36) means that the orthogonal projector on $\hp$ is given by
$\po$:
$$\hp=\po H \eqno (6.39)$$
The operator $N$ can be inverted on the orthogonal complement of its kernel:
$$\ns\=0P_0+\sum_{n=1}^\infty n^{-1}P_n\,\,,\,\,\ns N=N\ns=(1-\pp)\eqno(6.40)$$
Since $Q$ and $Q^+$ are $P_+^\uparrow$ invariant, so are $N$, $\ns$, and $P_n$:
$$U(a,\Lambda)\left\{N;\ns;P_n\right\}U(a,\Lambda)^{-1}=
\left\{N;\ns;P_n\right\} \eqno (6.41)$$
We also note that $N$, $P_n$, and $\ns$ commute with $Q$ and $Q^+$:
$$[Q^{(+)},N]_-=[Q^{(+)},P_n]_-=[Q^{(+)},\ns]_-=0 \eqno (6.42)$$

We now follow again [8] and introduce the following subspaces of $H$:
$$H_K\={\rm kernel}\,Q\,\,,\,\, H_{K^+}\={\rm kernel}\,Q^+\,\,,\,\,
  H_R\={\rm range}\,Q\,\,,\,\, H_{R^+}\={\rm range}\,Q^+ \eqno(6.43)$$
Let us study the relations between these spaces and $\hp$. Let $\Uao\in\hp$. By
$$0=\left(\Uao,\{Q,Q^+\}_+\Uao\right)=||Q^+\Uao||^2+||Q\Uao||^2 \eqno (6.44)$$
we find
$$\hp=H_K\cap H_{K^+} \eqno (6.45)$$
Since $Q^2=(Q^+)^2=0$ we have
$$H_R\subseteq H_K\,\,,\,\,H_{R^+}\subseteq H_{K^+} \eqno(6.46)$$
Let $\Uak\in \hk$, $\Ubrk\in\hrk$. Since
$$(\Uak,\Ubrk)=(\Uak,Q^+\Ub)=(Q\Uak,\Ub)=0 \eqno(6.47)$$
$\hk$ and $\hrk$ are orthogonal to each other, and replacing $Q$ by $Q^+$ shows
that the same holds true for $\hr$ and $\hkk$:
$$\hk\bot\hrk\,\,,\,\,\hr\bot\hkk \eqno(6.48)$$
Combining this with (6.46) gives
$$\hr\bot\hrk \eqno(6.49)$$
while (6.45) now implies
$$\hp\bot\hr\,\,,\,\,\hp\bot\hrk \eqno(6.50)$$
We conclude that the three spaces $\hp$, $\hr$, and $\hrk$ are all mutually
orthogonal. Let now $\Ua\in H$. Then we can write
$$\Ua=\po\Ua+(1-\po)\Ua=\po\Ua+N\ns\Ua=\po\Ua+\zw QQ^+\ns\Ua+\zw Q^+Q\ns\Ua
\=\Uao+\Uar+\Uark \eqno (6.51)$$
where $\Uao\in\hp$, $\Uar\in\hr$, and $\Uark\in\hrk$. This shows that the
Hilbert space $H$ is the direct orthogonal sum of the three spaces $\hp$,
$\hr$, and $\hrk$:
$$H=\hp\os\hr\os\hrk \eqno(6.52)$$
Since the first two of this spaces are subspaces of $\hk$ and the third is
orthogonal to it we can also write:
$$H=\hk\os\hrk\,\,,\,\,\hk=\hp\os\hr \eqno(6.53)$$
i.e the physical Hilbert space is also given by
$$\hp=\hk\om\hr\eqno(6.54)$$
Moreover, since $\hp$ and $\hrk$ are subspaces of $\hkk$ and this space is
orthogonal to $\hr$ one can also write
$$H=\hkk\os\hr\,\,,\,\,\hkk=\hp\os\hrk \eqno(6.55)$$
i.e. we get one more characterization of $\hp$ as
$$\hp=\hkk\om\hrk \eqno(6.56)$$
The orthogonal decompositions above were already given in [8], and, in the
specific context of massless Yang-Mills theories, in [4]. We denote the
orthogonal projections on $\{\hp;\hk;\hr;\hkk\hrk\}$ by
$\{\po;\pk;\pr;\pkk;\prk\}$ and the vectors in these spaces by
$\{\Uao;\Uak;\Uar;\Uakk;\Uark\}$. From the preceding equations we find:
$$\po\pr=\po\prk=\pr\prk=0\,\,,\,\,\po+\pr+\prk=1\,\,,\,\, \po=\po^+\,,\,
\pr=\pr^+\,,\,\prk=\prk^+\eqno(6.57)$$
$$\pr=\zw QQ^+\ns\,\,,\,\,\prk=\zw Q^+Q\ns \eqno(6.58)$$

Let us now study the strucure of some important operators with respect to
the orthogonal decomposition (6.52). The operators $Q$ and $Q^+$ map  the
complements of their kernels onto their range. This gives:
$$Q=\pr Q\prk\,\,,\,\, Q^+=\prk Q^+\pr \eqno(6.59)$$
Then (6.37) implies that the decomposition of $N$ and $\ns$ are given by
$$N^{(\sim 1)}=\pr N^{(\sim 1)}\pr+\prk N^{(\sim 1)}\prk \eqno(6.60)$$
Let now $I$ be a gauge invariant operator, i.e. $d_QI=0$. Then $\hk$ and
$\hr$ are stable under the action of $I$, that is
$$I\pk=\pk I\pk\,\,,\,\,I\pr=\pr I\pr \eqno(6.61)$$
Thus we get
$$I=\po I\po+\po I\prk+\pr I\po+\pr I\pr+\pr I\prk+\prk I\prk \eqno(6.62)$$
Next we use the pseudo-hermiticity of $Q$ and $Q^K$ to get information
about the Krein operator $J$. Let $\Uak\in\hk$, $\Ubr=Q\Uc\in\hr$. Then we have
$$(\Uak,J\Ubr)=<\Uak,Q\Uc>=<Q\Uak,\Uc>=0 \eqno(6.63)$$
This means:
$$\pk J\pr=0 \eqno(6.64)$$
Taking the adjoint gives:
$$\pr J\pk=0 \eqno(6.65)$$
Using $Q^+$ instead of $Q$ in the argument above gives
$$\pkk J\prk=\prk J\pkk=0 \eqno(6.66)$$
A direct inspection of $J$ in (6.31) gives the additional information that $J$
agrees on $\hp$ with the unit operator:
$$\po J\po=\po \eqno(6.67)$$
The last four equations are summarized in
$$J=\po J\po+\pr J\prk+\prk J\pr \eqno(6.68)$$

The second of eqs. (6.53) means that $\hk$ can be interpreted as a linear
fiber bundle: $\hp$ is the base space and the fibers are the elements of
$\hr$. Eqs. (6.64,6.65) show that the fibers are pseudo-orthogonal to any
vector in $\hk$. Moreover, writing $\Uak=\Uao+\Uar$ according to the orthogonal
decomposition (6.53) gives
$$<\Uak,\Ubk>=(\Uao,\Ubo) \eqno(6.69)$$
This shows that that the form $<,>$ agrees on $\hp$ with the positive form
$(,)$, that it is positive semidefinite in $\hk$, and that its
kernel as a quadratic form in this space are the fibers:
$$\hr={\rm kernel}\,<,>_K \eqno(6.70)$$
where $<,>_K$ means the restriction of the form $<,>$ to $\hk$.
So we get another expression for$\hp$:
$$\hp=\hk\om{\rm kernel}\, <,>_K \eqno (6.72)$$
The form $<,>_K$ is constant along the fibers in both arguments separately:
$$<\Uak+\Uar,\Ubk+\Ubr>_K=<\Uak,\Ubk>_K \eqno(6.73)$$
and the same holds true for the matrix elements of any gauge invariant
operator $I$ with respect to this form:
$$<\Uak+\Uar,I\left(\Ubk+\Ubr\right)>_K=<\Uak,I\Ubk>_K \eqno(6.73)$$
This allows to choose any linear cross section $H_S$ in $\hk$, i.e. any
subspace of $H$ which is a (pseudo-orthogonal but generally not orthogonal)
complement of $\hr$ (in $\hk$) as a realization of the physical Hilbert space.
The scalar product in $H_S$ is the restriction of the form  $<,>$ to this
space, and there it is positive definite.
All this spaces are unitarily equivalent, and the
matrix elements of gauge invariant operators do not depend on the section
chosen. So one might also consider the equivalence class of all this spaces
and that is what is usually done in the literature [7]. We prefer to use $\hp$
as a concrete realization of the physical Hilbert space since it is the only
section which is orthogonal to the fibers and which allows for a simple
interpretation of the quanta of the elementary fields as physical or
unphysical particles. The projections onto the sections along the fibers are
also called gauge transformations. For $\hp$, and only for it, they agree with
the orthogonal projection.

We now consider again operators $A$, $B$, $C\cdots$ over $H$. We define the
orthogonal projection of $A$ on $\hp$: $A_0$ by
$$A_0\=\po A\po \eqno(6.74)$$
$A_0$ is still an operator from $H$ to$H$. It is zero on the orthogonal
complement of $\hp$: $\hp^\bot$. Since this zero is certainly not very
interesting we define $\ap$ to be the restriction of $A_0$ to $\hp$:
$$\ap\=\left(A_0\right)_{\downarrow\hp} \eqno(6.75)$$
The map $A\longrightarrow\ap$ is certainly linear:
$$(\al A)_\p=\al\ap\,\,,\,\,(A+B)_\p=\ap+\bp \eqno(6.76)$$
More interesting is the projection of the product of two operators. We
calculate:
$$\po AB\po=\po A(\po+\pr+\prk)B\po=\po A\po\po B\po+\po A\zw QQ^+\ns B\po+
\po A\zw\ns Q^+QB\po \eqno(6.77)$$
Let us concentrate on the second summand: X. Since $\po Q$=0 we can replace
$AQ$ by $\{A,Q\}_\pm$. We take the anticommutator if $A$ is fermionic and the
commutator if it is bosonic. This gives
$$X=\po\zw\{A,Q\}_\pm Q^+\ns B\po \eqno(6.78)$$
Now we use $\po Q^+=0$ to replace that by
$$X=\po\zw\left\{\{A,Q\}_\pm,Q^+\right\}_\mp\ns B\po \eqno(6.79)$$
And finally we use $\po\ns=0$ to write
$$X=\po\zw\left[\left\{\{A,Q\}_\pm,Q^+\right\}_\mp,\ns\right]_-B\po
\eqno(6.80)$$
There are always two commutators and one anticommutator in this expression.
Thus it can be uniquely written as
$$X=\po(\T A)B\po \eqno(6.81)$$
where the {\sl triple variation} $\T$ is defined by
$$\T\=\zw \de_{(\ns)}d_{(Q^+)}d_Q \eqno(6.82)$$
Here $\de_{(\ns)}$ is the derivation induced by $\ns$, and $d_Q$ and
$d_{(Q^+)}$ are the antiderivations induced by $Q$ and $Q^+$, repectively
(see chapter 1). Note that the triple variation of gauge invariant operators
vanishes. In the same way the third summand in (6.77): $Y$ is written as
$$Y=\po A(\T B)\po \eqno(6.83)$$
We thus have found the important {\sl projection formula}:
$$\ap\bp=(AB)_\p-\left\{(\T A)B+A(\T B)\right\}_\p \eqno(6.84)$$
This implies

\vskip 0.2cm

{\bigtype Theorem I:} The product of the physical projections of two
operators with vanishing triple variation, especially of two gauge invariant
operators, is identical to the physical projection of their product. The
physical projection of a group (of an algebra) of operators with vanishing
triple variation, especially of gauge invariant operators, is a representation
of this group (algebra).

\vskip 0.2cm

Next we consider the physical projection of the pseudo-adjoint $A^K$ of an
operator $A$. So we have to study
$$\po A^K\po=\po JA^+J\po \eqno(6.85)$$
Now we use that (6.68) implies
$$\po J=J\po=\po J\po=\po \eqno(6.86)$$
to conclude:
$$\po A^K\po=\po A^+\po=(\po A\po)^+ \eqno(6.87)$$
which means
$$\left(A^K\right)_\p=\left(\ap\right)^+ \eqno(6.88)$$
Thus we have found

\vskip 0.2cm

{\bigtype Theorem II:} The physical projection of the pseudo-adjoint operator
is identical to the adjoint of the physical projection of this operator. The
physical projection of a pseudo-hermitean operator is hermitean.

\vskip 0.2cm

Combining the two theorems above gives:

\vskip 0.2cm

{\bigtype Theorem III:} The physical projection of a pseudo-unitary operator
with vanishing triple variation, especially of a pseudo-unitary gauge
invariant operator, is an unitary operator.

\vskip 0.2cm

Now we are well equipped to tackle unitarity of the physical $S$-matrix.
The interaction $T^{(1)}(x)$ constructed in the preceding chapters is
anti-pseudo-hermitean:
$$\left(T^{(1)}(x)\right)^K=-T^{(1)}(x) \eqno(6.89)$$
This guarantees the pseudo-unitarity of $S[g]$ [11,12,15]:
$$S[g]S^K[g]=1 \eqno(6.90)$$
This is [11] equivalent to
$$\sum_{I\oplus J=N}T(I)T^K(J)=0,\,\,\,\forall N\neq\emptyset \eqno(6.91)$$
Here $T(I)$ means $T^{(r)}(x_{i_1},\cdots, x_{i_r})$, $T^K(J)$ means
$\left(T^{(s)}(x_{j_1},\cdots,x_{j_s})\right)^K$, where $r+s=n$, and the sum
$\Sigma$ runs over all direct decompositions of the set
$N=\{1,\cdots,n\}$ into two subsets $I=\{i_1,\cdots,i_r\}$ and
$J=\{j_1,\cdots,j_s\}$.
Let us now {\sl assume} that gauge invariance holds true in all orders, i.e.
we have
$$d_QT^{(n)}=\sum_{k=1}^{4n}\d^kT^{(n)}_k \eqno(6.92)$$
Taking the pseudo-conjugate of this equation gives:
$$d_Q\left(T^{(n)}\right)=-\sum_{k=1}^{4n}\d^k\left(T^{(n)}_k\right)^K
\eqno(6.93)$$
Then (6.84,6.88) and (6.90-6.92) imply
$$\sum_{I\oplus J=X}T_\p(I)\left(T_\p(J)\right)^+=
\sum_{k=1}^{4n}\d^kW^{(n)}_k(X)\,\,,$$
$$W^{(n)}_k(X)\=\zw\sum_{ I\oplus J=X}
\left\{-\theta_I(k)\left(\de_{(\ns)}d_{(Q^+)}T_k(I)\right)T^K(J)
+\theta_J(k)T(I)\left(\de_{(\ns)}d_{(Q^+)}T^K_k(J)\right)\right\}_\p
\eqno(6.94)$$
where
$$\theta_I(k)\=\cases{1,&if $k\in I$\cr 0,&if $k\notin I$\cr} \eqno(6.95)$$
This is the exact {\sl perturbative, pre-adiabatic} expression for the
unitarity of the physical $S$-matrix.
{\sl If} the adiabtic limit:
$$\sp=\lim_{g\to1}\left(S[g]\right)_\p \eqno(6.96)$$
exists and has the same analytic properties as in the saclar theory discussed
in [13], and {\sl if} the boundary terms $\int\,\d^kW^{(n)}_k$ vanish, (6.94)
will imply
$$\sp\sp^K=1 \eqno(6.97)$$
We note, however, that this adiabatic limit may have additional subtleties
if the heavy gauge particles are coupled to light matter, since then the gauge
fields become unstable, i.e are not really asymptotic fields.

Let us end this long chapter with a short discussion of
$P_+^\uparrow$-invariance. The remark after (6.32) and eq. (6.41) immediately
imply that $U(a,\Lambda)_\p$ is an unitary representation of $P_+^\uparrow$,
while the $P_+^\uparrow$-invariance of $S[g]$:
$$U(a,\Lambda)S[g]U(a,\Lambda)^{-1}=S[g_{a,\Lambda}]\,\,,\,\,
g_{a,\Lambda}(x)\=g\left(\Lambda^{-1}(x-a)\right) \eqno(6.98)$$
and (6.41) lead direct to
$$U(a,\Lambda)S[g]_\p U(a,\Lambda)^{-1}=S[g_{a,\Lambda}]_\p \eqno(6.99)$$
which is the $P_+^\uparrow$-invariance of the physical $S$-matrix.

The situation is different in the massless case where $J$ is not
$P_+^\uparrow$-invariant. However, the three theorems above and the projection
formula (6.84) hold true in this case, too. Since $Q$ is
$P_+^\uparrow$-invariant also in the massless case, the theorems show that
$U(a,\Lambda)_\p$ is indeed an unitary representation of $P_+^\uparrow$, while
the $P_+^\uparrow$-invariance of $S[g]$ and (6.84,6.88) implie that (6.99)
is violated only by boundary terms. The latter {\sl should} vanish in physical
quantities like cross sections, for example.

\vskip 1cm

{\bigtype\bf 7. Discussion}

\vskip 0.2cm

We first have shown that the interaction of massless Yang-Mills fields studied
in [1-4] admits generalizations. Then we have constructed gauge invariant
first order interactions between massive Yang-Mills fields, scalar gauge
fields, and matter fields. We have shown how invariance under ghost charge
conjugation fixes the interaction uniquely.
The scalar gauge fields had to be introduced for a
pure algebraic reason, i.e. to have $Q^2$=0. Moreover, we have proven that
gauge invariance implies unitarity of the physical $S$-matrix and that the
latter is Poincar\'e invariant. However, gauge invariance were only shown to
hold true in the first order. This is certainly not enough. Before we can
claim to have a completely consistent theory we have to proof gauge invariance
in all orders, i.e. the absence of anomalies. We intend to tackle this
labourious though important task in future publications.

The analysis of unitarity has shown that {\sl all} scalar gauge fields are
unphysical. This means that our (pure) gauge theory is not a Higgs-model.
One could, of course, introduce a physical scalar particle in
the matter sector of the theory. But there seems to be no logical reason
for doing so (unless one finds out that this would be the only way to avoid
anomalies) since, in contrast to the Higgs model, the matter sector and the
gauge sector are independent structures in our theory.
Our theory has structural similarity to Stueckelberg type gauge theories.
There [20] all scalar gauge fields are unphysical, too. We remark,
however, that our theory is not identical to the Stueckelberg models discussed
in [20], it differs from them in the detailed structure of the propagators
and couplings. Stueckelberg type theories are most often discussed under the
aspect of classical gauge symmetry and gauge independence while BRS-invariance
is only a derived concept. We have followed a completely different route and
have made (free) BRS-invariance a first principle. This seems appropriate
since it is exactly this invariance which implies unitarity. All Stueckelberg
models studied in [20] where either nonrenormalizable or nonunitary. This
was shown by direct calculation, but the reason why remained a mystery. Our
model is renormalizable by construction, and unitarity follows from gauge
invariance. So, {\sl should} it fail to be consistent, we know at least why:
There have to be anomalies. One could then study these anomalies to learn how
to avoid them.

To proof unitarity of the physical $S$-matrix we have extended Razumov\'s and
Rybkin\'s investigation of quadratic BRS systems [8] by giving the important
projection formula (6.84). This formula shows that the triple variation $\T$
is more inportant for the algebraic structure of the theory than the gauge
variation is. Indeed, it follows from our calculations in the last chapter
that a theory which is not gauge-invariant but is invariant under the triple
variation $\T$ would still have an unitary physical $S$-matrix! Thus one might
call anomalies of gauge invariance which do not violate triple invariance
{\sl weak anomalies} and anomalies which do violate it {\sl strong anomalies}.
Well, it {\sl} could happen that all {\sl known anomalies} are strong ones,
but this is far from obvious. A careful algeraic analysis of anomalies with
respect to the triple variation is certainly worth doing, and we plan to
investigate this interesting structure in the future.

Another thing which one has to do before our theory can make contact with
``physics'' is the extension to the nonsimple group $G={\rm U}(2)$ and an
analysis of global $G$-breaking which allows for different masses of the
various gauge fields.

Let us summarize: A direct algebraic analysis of massive {\sl quantized}
Yang-Mills theories done in the framework of causal perturbation theory
and free of any classical notions as spontaneuos symmetry breaking has
naturally led to the construction of a Higgs free model of massive gauge fields
and has revealed new interesting algebraic structures. Further investigations
remain to be done.

\vskip 1cm

{\bigtype\bf Acknowledgments - }I would like to thank: Ivo Schorn for providing
me with an excellent \TeX nical enviroment and for his constant and patient
advise how to use it. Dr. Michael Duetsch for carefully proofreading the
manuscript. The Swiss National Science Foundation for financial support.
And the Dublin Institute for Advanced Studies and the University of
Zuerich for warm hospitality.

\vskip 1cm

{\bigtype{\bf References}}

\vskip 0.2cm

\noindent 1. M. Duetsch, T. Hurth, F. Krahe, and G. Scharf,
{\it Nuovo Cimento A} {\bf 106} (1993), 1029.

\noindent 2. M. Duetsch, T. Hurth, F. Krahe, and G. Scharf,
{\it Nuovo Cimento A} {\bf 107} (1994), 375.

\noindent 3. M. Duetsch, T. Hurth, and G. Scharf,
{\it Nuovo Cimento A} {\bf 108} (1995), 679.

\noindent 4. M. Duetsch, T. Hurth, and G. Scharf,
{\it Nuovo Cimento A} {\bf 108} (1995), 737.

\noindent 5. F. Krahe, ``On The Algebra of Gauge Fields'',
preprint DIAS-STP-95-02, hep-th-9502097.

\noindent 6. W. H. Greub,
`` Linear Algebra'',
3rd ed., Springer, New York, 1967.

\noindent 7. N. Nakanishi and I. Ojima,
`` Covariant Operator Formalism Of Gauge Theories And
Quantum Gravity'',

\noindent\hphantom {7. }World Scientific, Singapore, 1990.

\noindent 8. A. V. Razumov and G. N. Rybkin,
{\it Nucl. Phys. B} {\bf 332}, (1990), 209.

\noindent 9. N. N. Bogolubov and D. V. Shirkov,
`` Introduction To The Theory Of Quantized
Fields'',

\noindent\hphantom {9. }Interscience, New York, 1959.

\noindent 10. N. N. Bogolubov and D. V. Shirkov,
`` Quantum Fields,''
Benjamin-Cummings, Reading (Mass), 1983.

\noindent 11. H. Epstein and  V. Glaser,
{\it Ann. Inst. H. Poincar\'e} {\bf 19}, (1973), 211.

\noindent 12. G. Scharf,
`` Finite QED'', Springer, Berlin, 1989.

\noindent 13. H. Epstein and  V. Glaser,
`` Adiabatic Limit in Peruturbation Theory'',
{\it in} ``Renormalization Theory'',

\noindent\hphantom {13. }(G. Velo and A. S. Wightmann, Eds.),
pp. 269-297, D. Reidel, Dordrecht (NL), 1976.

\noindent 14. P. Blanchard and R. S\'en\'eor,
{\it Ann. Inst. H. Poincar\'e} {\bf 23}, (1975), 147.

\noindent 15. M. Duetsch, ``On Gauge Invariance of Yang-Mills Theories with
Matter Fields'',
preprint ZU-TH-10-95

\noindent 16. J. Bognar,
`` Indefinite Inner Product Spaces'',
Springer, Berlin, 1974.

\noindent 17. N. N. Bogolubov, A. A.Logunov, A. I. Oksak, and I. I. Todorov,

\noindent\hphantom {17. } ``General Principles Of Quantum Field Theory'',
Kluwer, Dordrecht (NL), 1990.

\noindent 18. S. S. Schweber,
``Relativistic Quantum Field Theory'',
Harper and Row, New York, 1961.

\noindent 19. F. A. Berezin,
``The Method Of Second Quantization'',
Academic Press, New York, 1966.

\noindent 20. R. Delbourgo, S. Twisk, and G. Thompson,
{\it Int. J. Mod. Phys. A} {\bf 3}, (1988), 435.

\bye